\documentclass[preprint,aps]{revtex4}
\usepackage{amsmath}
\usepackage{amssymb}
\usepackage{graphicx}

\begin{document}

\title{Scaling behavior of a one-dimensional correlated\\
disordered electronic\\
System}
\author{Ibrahim Avgin}
\affiliation{Department of Electrical and Electronics\\
Engineering, Ege University,\\
Bornova 35100, Izmir, Turkey}
\email{ibrahim@eng.ege.edu.tr}
\date{\today}

\begin{abstract}
A one-dimensional diagonal tight binding electronic system with correlated
disorder is investigated. The correlation of the random potential is
exponentially decaying with distance and its correlation length diverges as
the concentration of "wrong sign" approaches to $1$ or $0$. The correlated
random number sequence can be generated easily with a binary sequence
similar to that of a one-dimensional spin glass system. The localization
length (LL) and the integrated density of states (IDOS) for long chains are
computed. A comparison with numerical results is made with the recently
developed scaling technique results. The Coherent Potential Approximation
(CPA) is also adopted to obtain scaling functions for both the LL and the
IDOS. We confirmed that the scaling functions show a crossover near the band
edge and establish their relation to the concentration. For concentrations
near to $0$ or $1$ (longer correlation length case), the scaling behavior is
followed only for a very limited range of the potential strengths.
\end{abstract}

\maketitle


Theoretical interest in disordered chains remains strong. Recent
investigations of the correlated disordered in random or chaotic arrays
revealed surprising results such as a possible breakdown of the Anderson criterion
for the localization \cite{hy04,lyra02,lyra98,izra99,deyc03}. Various
fields have made use of the results obtained from the study of the
one-dimensional models. Random microwave transmission in a single-mode wave
guide experiment \cite{ul}, transport studies with GaAs-AlGaAs random dimer
super lattice systems \cite{loc}, and the photonic band-gap structures \cite%
{bay} are a few examples. Recently a new renormalization technique \cite%
{rus98,rus02} has been introduced to study the scaling behavior of the well
known tight binding chain with long range correlated disorder. The authors
found that the Localization Length (LL) shows scaling behavior and a cross
over near the band edge. The behavior of disordered magnetic or electronic
chains \cite{ishii} (using proper transformations \cite{avgin93,pimentel89})
can be described mathematically by a tri-diagonal tight binding model given
by

\begin{equation}  \label{tb}
(E -\xi_{n}V)\psi_{n}=\psi_{n+1}+\psi_{n-1},
\end{equation}

\noindent where $E$ is energy, $V$ is the strength of the random potential
with its correlated random sign $\xi_{n}$. $n$ is the site index.

In this paper, we study the scaling properties of a particular form of
correlated disorder that is associated with spin glass chains \cite%
{pimentel89,avgin93,avgin02,boukahil89}. The random sign has the relation $%
\xi _{n}=\xi _{n-1}x_{n}$ with the following distribution $%
P(x_{n})=(1-c)\delta (x_{n}-1)+c\delta (x_{n}+1)$ where the $x_{n}$ are
uncorrelated between different sites and $c$ is the concentration of "wrong
signs". Clearly $\xi _{n}$ is exponentially correlated, \emph{i.}$\,$\emph{e.%
}, $\langle \xi _{n}\xi _{m}\rangle =(1-2c)^{|m-n|}$ where one can define a
correlation length \cite{vulp89} $l(c)=-1/ln|1-2c|$. For $c>0.5$ the
ordering is of "antiferromagnetic type" and $c=0.5$ is uncorrelated case
since the correlation length is zero. As seen in Fig.~1, for $c$ approaching 
$0$ or $1$ the correlation length diverges.

Previous studies of the disordered chain problem showed that the electronic
wave function decays exponentially with a distance \cite{thouless}
characterized by the real part of the Lyapunov exponent \cite{derrida84}
whose imaginary part is also related to the Integrated Density of States
(IDOS) \cite{derrida84}. In our work numerical calculations are carried out
for chains of $10^{8}$ sites using the negative eigenvalue counting
technique introduced by Dean \cite{dean} providing direct computation of the
Lyapunov exponent.

Recently the scaling properties arising from long range correlated disorder 
\cite{rus98,rus02} were investigated using the renormalization approach. The
authors have argued that their analyses work close to the band edge where
the characteristic wavelength diverges since neighboring lattice sites move
as blocks. Here we adopt their results for the correlated case discussed
above.

The complex Laypunov exponent \cite{derrida84} is defined as $\Gamma(E,V,c)
= \lim_{N\to\infty}\frac{1}{N} \ln \frac{\psi_N}{\psi_0}$. The space
decimation procedure \cite{rus98} leads to redefining the random potential
by its mean value over the block, \emph{i.}$\,$\emph{e.}, $%
V_N=\sum_{n=1}^NV\xi_n$.  The square of this scaled potential has finite value
proportional to $N$ with increasing $N$ for all concentrations
as displayed in Fig.~1.  For $c>0.5$, the summation has alternating values
particularly persistent for $c \rightarrow 1$ 
but eventually reaches to a limiting value
discussed below for larger $N$ (see Fig.~1). 
For uncorrelated case exact scaling results 
exist for the $LL$ and the $IDOS$ 
defined via the Lyapunov exponent. Near the band edges ($E \to 2$ and
$V \to 0$) the Lyapunov exponent displays 
a scaling law of the form \cite{derrida84}
$\Gamma \approx <V^2>^{1/3}h(\frac {2-E}{<V^2>^{2/3}})$ where
$h$ explicitly known scaling function. For the correlated
case, it was assumed that the Lyapunov exponent \cite{rus98} 
has similar power-law relation to the second moment of the redefined
random potential  (note that the first moment is zero) 
$<V_N^2>$ which can be written for our case as:

\begin{equation}
<V_N^2>=\frac{V_N^2}{N}=\frac{1-c}{c}V^2.  \label{vsq}
\end{equation}

\noindent Detailed account of this factor is given in Ref.\cite{avgin98,avgin93} 
This factor (in fact its $1/3$ power) $\frac{1-c}{c}$ is shown in
Fig.~1 and it plays a role in the scaling behavior discussed below. Unlike
the correlation length it diverges as $c\rightarrow 0$ while it remains
finite as $c\rightarrow 1$.

\begin{figure}[tbp]
\centering
\includegraphics[angle=270, width=18cm]{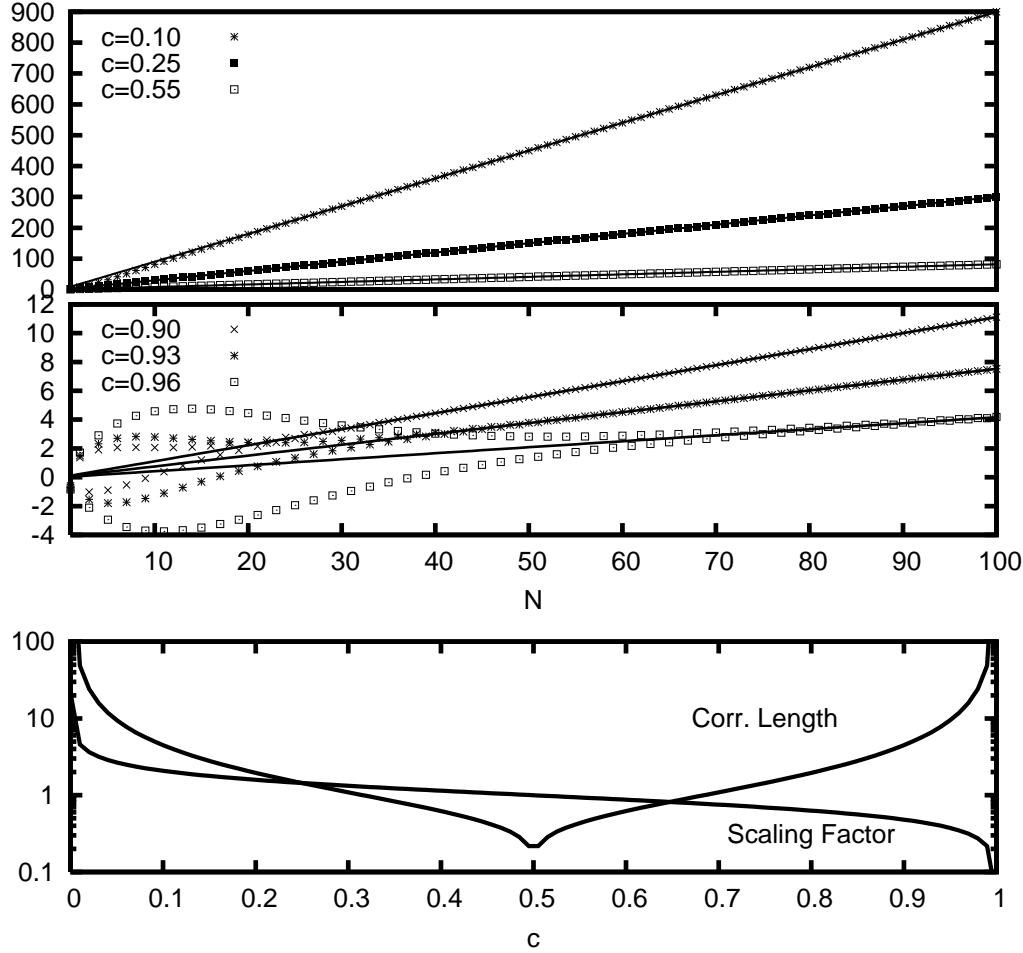}
\caption{ The lower panel is a plot of the scaling factor 
$(\frac {1-c}{c})^{1/3}$ and the correlation length $l(c)$ versus $c$. 
$l(c)$ is symmetric about $c=0.5$ and zero at this concentration. The
square of the potential $V_{N}^2$ versus the number of cites are
presented at the upper panels with respective concentrations.  The lines
are $V_{N}^2=\frac {1-c}{c}N$ obtained for a large $N$
and the points are calculated by a finite summation of $V_{N}^2$.}
\label{fig:scl}
\end{figure}

At the band edge $E=2$, according to the analysis of Ref.\cite{rus98} 
the space decimation for blocks of b sites 
results in the following scaling: the strength of the potential must
hold (to preserve the form of the Eq.~(\ref{tb}) 
$ V \rightarrow V_{b}=bV$ (see Eq.(7-11) of Ref.\cite{rus98}), hence
the second moment transforms 
$<V_{N_b}^{2}> = b\frac {1-c}{c} (bV)^2=b^{3}<V_{N}^{2}>$. 
This corresponds to their \cite{rus98} case of $\gamma = 1$. 
Thus The Lyapunov exponent at the band edge can be written in a form 
\cite{derrida84,rus98} 
$\Gamma (E=2,V,c)=(<V_{N}^{2}>)^{1/3}=(V^{2}\frac{1-c}{c})^{1/3}$.
The power of the Lyapunov exponent is the same as the uncorrelated case 
\cite{derrida84} but the coefficient is different. 
The factor involving concentration
was recovered before using phenomelogical arguments 
in the context of the spin glass chain \cite%
{avgin93,avgin98}. 
The exact results for uncorrelated case $c=0.5$ are well known \cite%
{derrida84} given by $IDOS=-{\frac{1}{\pi }}\Im \Gamma =0.159V^{2/3}$ and $%
\frac{1}{LL}=\Re \Gamma =0.289V^{2/3}$. In an earlier work we showed that
the IDOS and the LL \cite{avgin93,avgin98,avgin02} results for various
concentrations scaled with a similar factor where data collapsed well to the
exact calculation at $c=0.5$.

For uncorrelated case it is known that there is a cross over limit when $%
w\equiv 2-E$ is positive and very small \cite{rus02,derrida84}. In this
limit, the Lyapunov exponent can be written in the form

\begin{equation}
\Gamma (w,V,c) \sim (V^2 \frac{1-c}{c})^{1/3}F(X)
\end{equation}%
\ 

\noindent where $X\equiv w(V^{2}\frac{1-c}{c})^{-2/3}$ and $F(X)$ is the
scaling function. From the previous studies it was found \cite%
{derrida84,rus02} that the scaling function has different behaviors for $X<<1
$ and $X>>1$, hence there is a cross over between these two limits. It was
argued \cite{rus02} that in this system the cross over is a consequence of
the competition between the characteristic length $\lambda $ of the wave
function and the LL and the two asymptotic regimes can be obtained by the
dominance of the LL or $\lambda $. When $\lambda >>LL$ where $w\rightarrow 0$
the wave function is independent of $\lambda $ since it would decay over a
distance of $\lambda $ and the LL does not deviate much from its value at
the band edge so that $F(x)\rightarrow constant$. In the other limit $%
LL>>\lambda $ where the wave function would have considerable oscillations
before decaying, the LL has a different functional dependence on the
strength of the random potential \cite{rus02,derrida84}, \emph{i.}$\,$\emph{%
e.}, $LL\sim <V^{2}>$ since $w$ dominates the dynamics yielding $%
F(X)\rightarrow X$.

Insight can be gained if we adopt the Coherent Potential Approximation (CPA)
results here. The CPA self energy or the coherent potential has important
connections to the system's dynamics which for uncorrelated disorder \cite%
{avgin96,avgin2} can be written as $V_{c}\simeq (\frac{<V^{2}>}{2i})^{2/3}$.
For correlated random potentials we can substitute here the second moment of
the potential given in Eq.~(\ref{vsq}) (so that the correlations between
different sites are included) then the coherent potential takes this form $%
V_{c}=(\frac{1-c}{c}\frac{V^{2}}{2i})^{2/3}$. The dispersion relation can be
obtained from the poles of the configurationally averaged $k$ dependent
Green function \cite{avgin96} that $E-\Re V_{c}=2cosk$ where $k$ is a wave
vector. For small $k$, it can be reformulated as $k^{2}\simeq \Re V_{c}+w$.
Away from the band edge the LL was calculated before \cite%
{rus02,derrida84,avgin2} and can be expressed as $LL=\frac{8{\sin }^{2}k}{%
<V^{2}>}\simeq \frac{8k^{2}}{<V^{2}>}$. Recalling the fact that in one
dimension $k$ is proportional to the IDOS, both the LL and the IDOS can be
rearranged yielding the scaling functions respectively $F(x)=LL(\frac{1-c}{c}%
V^{2})^{1/3}=8(2^{-5/3}+X)$ and $G(X)=IDOS(\frac{1-c}{c}V^{2})^{-1/3}=
{\pi}^{-1}\sqrt{2^{-5/3}+X}$ where $G(X)$ has different asymptotic 
behavior when $X>>1$, $G(X)\rightarrow \sqrt{X}$.

We present our numerical and the CPA results below. We have calculated the
scaling functions $F(X)$ and $G(X)$ for chains of $10^{8}$ sites using the
negative eigenvalue counting technique. As shown in Fig.~1 (semi-logarithmic
plot) the correlation length goes to zero as $c\rightarrow 0.5$ and
increases with increasing $|c-0.5|$ with a rapid increase developing near $%
c\rightarrow 1,0$. The results presented in Fig.~2 are for shorter
correlation lengths where  $c=0.1,0.25,0.5,0.75,0.9$ and for $w=10^{-4}$ and 
$10^{-3}$. The data clearly indicate that there is a scaling function for
both the LL and IDOS. The CPA under estimates the scaling function for the
LL but reproduces the IDOS results rather well as shown in figures presented
in this section. We have found that for $w>10^{-4}$ the scaling functions
some what deviate from the expected scaling behavior near the lower values
of $X$. However for $w<10^{-4}$ they follow the predicted results.

\begin{figure}[tbp]
\centering
\includegraphics[angle=270,width=12cm]{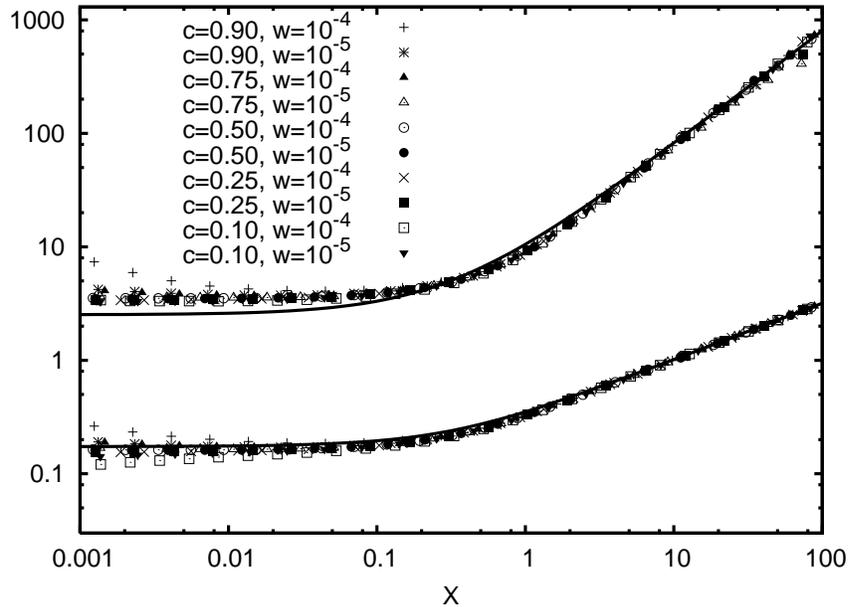}
\caption{ plot of $F(X)$ (top) and $G(X)$ (bottom) as function of the
scaling variable $X \equiv w (V^2 \frac{1-c}{c})^{-2/3}$ strength $V$. The
lines are the CPA results.  The data displayed for the concentrations $%
c=0.1,0.25,0.5,0.75,0.9$ and the two energies $w=2-E=10^{-4}$ and $10^{-5}$.}
\label{fig:sc1}
\end{figure}

\begin{figure}[tbp]
\centering
\includegraphics[angle=270,width=12cm]{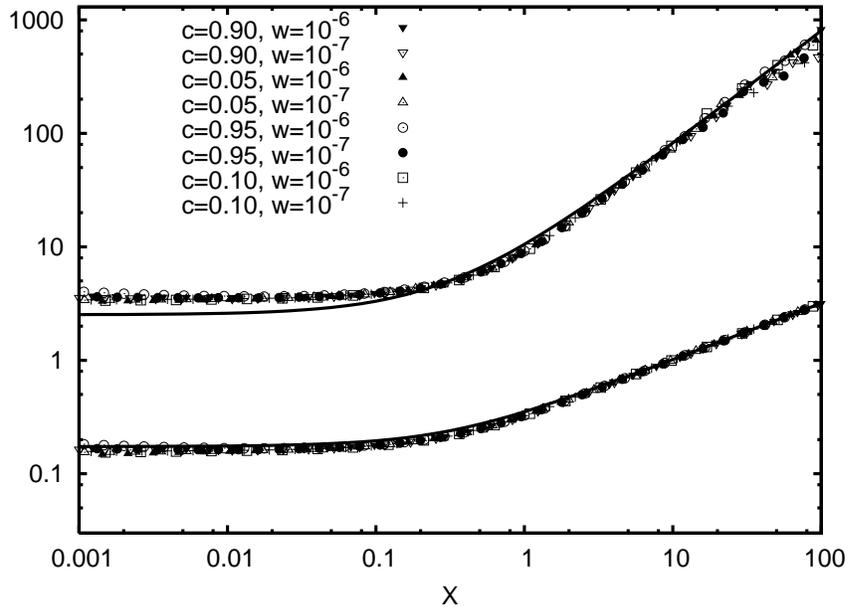}
\caption{ plot of $F(X)$ (top) and $G(X)$ (bottom) as function of the
scaling variable $X \equiv w (V^2 \frac{1-c}{c})^{-2/3}$ strength $V$. The
lines are the CPA results.  The data displayed for the concentrations $%
c=0.1,0.05,0.9,0.95$ and the two energies $w=2-E=10^{-6}$ and $10^{-7}$.}
\label{fig:sc2}
\end{figure}

For concentrations near $0,1$ a region of rapidly increasing correlation
length, we produced two sets of data displayed in Fig.~3 and Fig.~4. In
Fig.~3 the scaling functions are displayed for concentrations $%
c=0.1,0.05,0.9,$and $0.95,$ where the rapid increase in the correlation
length starts. The scaling behavior is observed for only $w\geq 10^{-6}$.
Note that we obtained similar, even better scaling behavior for the
concentrations presented in Fig.~2 with $w\geq 10^{-6}$ values.

The scaling behavior is also clearly revealed in Fig.~4 where the deviations
near the lower limit are generally largest for $w=10^{-6}$. The
concentrations here are very close to $0,1$, as presented in figure caption,
so one has to go to very small values of $w$ (fixed $V$) to see the
scaling behavior. The shrinking of the scaling region in $w$ as $c \to 0,1$
is consistent with the behavior for $c = 0,1$, where the correlation 
length is infinite. The collapse of the data for different values of $c$ 
is a feature of the model that reflects the fact that both the 
second moment (Eq.~(\ref{vsq}) and the correlation 
length are functions of $c$. 
The IDOS results also deviate for this range of
concentrations. Thus as the correlation length is increased, the scaling
behavior can be observed only for smaller and smaller values of $w$;
however, we found that $w$ cannot be reduced as much as desired.  Hence 
after certain small $w$, deviations starts this time 
particularly at higher $X$ sides.  
Thus there is a trade-off in building the scaling variable $X$ since
it depends on the three parameters $w$, $c$ and $V$ and only
delicate combination of these parameters produces scaling and the
crossover behavior.  Stronger deviations presented in Fig.~3-4
at very small $X$ for the concentrations near $0,1$ show 
this sensitive dependence.  Because very small concentrations, the 
scaling factor becomes very small or very large, therefore the 
Lyapunov exponent must be computed in a region 
of a very small or a very large potential strengths at very small 
$w$ wherein the computational errors inevitable
or the scaling behavior of the Lyapunov exponent holds only very
limited potential strengths $V$. 

\begin{figure}[tbp]
\centering
\includegraphics[angle=270,width=12cm]{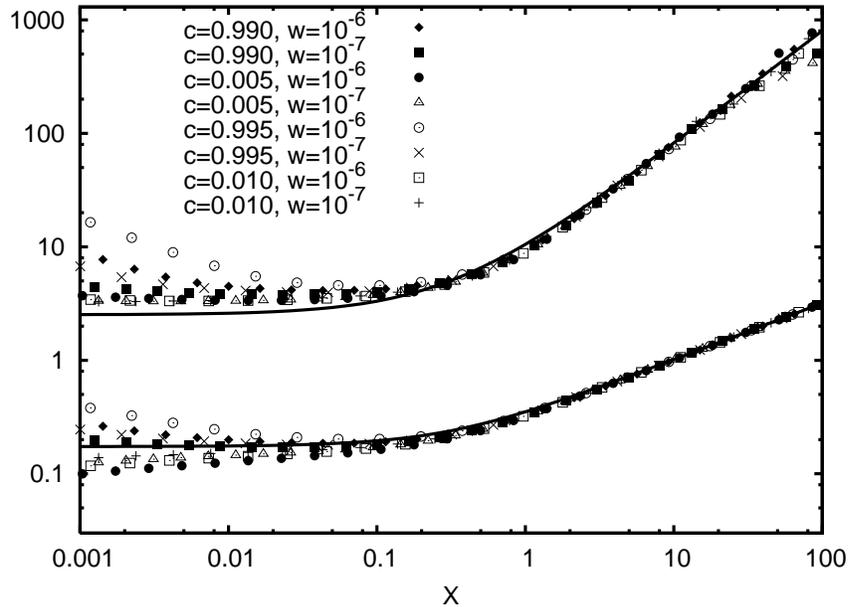}
\caption{ plot of $F(X)$ (top) and $G(X)$ (bottom) as function of the
scaling variable $X \equiv w (V^2 \frac{1-c}{c})^{-2/3}$ strength $V$. The
lines are the CPA results. The data displayed for the concentrations $%
c=0.01,0.005,0.99,0.995$ and the two energies $w=2-E=10^{-6}$ and $10^{-7}$.}
\label{fig:sc3}
\end{figure}

We have studied binary correlated disordered chains both analytically and
numerically. The LL and the IDOS are computed for various concentrations. We
found that they both showed a scaling behavior and a crossover. The behavior
of the scaling functions predicted by the renormalization group techniques 
\cite{rus02,rus98} is observed. The data calculated numerically for various
concentrations with increasing correlation length largely collapsed to
scaling functions belonging to the LL and IDOS separately. The scaling
factor as a function of concentration plays a key role in the scaling
behavior of the LL and IDOS data. The scaling behavior obtained here holds
for a more limited range of $w$ values than that obtained by Russ \emph{et}$%
\,$\emph{al.} \cite{rus02}. For instance we did not see any scaling behavior
when $w>10^{-4}$, whereas they presented the scaling behavior for $%
w=10^{-1}-10^{-5}$. For the scaling near the limiting concentrations, some
scattered data are obtained for lower values of the scaling variable $X$. We
have used scaling arguments and established the scaling functions for both
the LL and the IDOS using the CPA results. The CPA results reproduced data
for the ILL rather qualitatively but surprisingly for the IDOS it reproduced
the data rather well. This study revealed that if the second moment of the 
correlated random potential is calculated, then the CPA results can be
implemented as the way presented here even though the CPA was developed
mainly for noncorrelated random potentials.

We have benefited from discussions with Prof. D. L. Huber. This work is
partially sponsored by the Scientific end Technical Research Council of
Turkey (TUBITAK).


\end{document}